\documentclass[%
 reprint, %superscriptaddress, %groupedaddress, %unsortedaddress, %runinaddress, %frontmatterverbose, 
 amsmath,amssymb,
 aps,
 pra,
]{revtex4-2}

\usepackage{soul}%\st for strike through
\usepackage[normalem]{ulem}
\usepackage{xcolor}
\usepackage{graphicx}
\usepackage{hyperref}
\hypersetup{
    colorlinks=true,
    linkcolor=blue,
    citecolor=magenta,
    filecolor=magenta,      
    urlcolor=blue,
    pdftitle={}, %Title of the pdf file
    }
\newcommand{\xref}[1]{(\ref{#1})}
\newcommand{\be}{\begin{equation}}
\newcommand{\ee}{\end{equation}}
\newcommand{\bs}{\boldsymbol}
\newcommand{\mbf}{\mathbf}

\newcommand{\bx}{\mbf{x}}

\newcommand{\ebold}{\mbf{E}}
\newcommand{\bbold}{\mbf{B}}
\newcommand{\z}{\hat{\mbf{z}}}
\newcommand{\s}{\hat{\mbf{r}}}
\newcommand{\bphi}{\hat{\bs{\varphi}}}
\newcommand{\nablab}{\nabla_\bot}
\newcommand{\divbot}{\nablab \cdot}
\newcommand{\ep}{\bs{\mathcal{E}}}
\newcommand{\bp}{\bs{\mathcal{B}}}
\newcommand{\bb}{\mbf{b}}
\newcommand{\bbot}{\mbf{b}_\bot}

\begin{document}
%%%%%%%%%%%%%%%%%%%%%%%%%%%%%%%%%%%%%%%%

%\def\baselinestretch{1}\selectfont
\preprint{APS/123-QED}

\title{Probing emergent axion electrodynamics with waveguide apparatus}

\author{Andr\'e H. Gomes}
\email{andre.gomes@ufop.edu.br}
\affiliation{Departamento de F\'{i}sica, Universidade Federal de Ouro Preto, 35402-163,\\ Ouro Preto, Minas Gerais, Brazil}%

\author{Winder A. Moura-Melo}%
\email{winder@ufv.br}
\affiliation{Departamento de F\'{i}sica, Universidade Federal de Vi\c{c}osa, 36570-900,\\ Vi\c{c}osa, Minas Gerais, Brazil}%

%\date{\today}

\begin{abstract}
We demonstrate that a conventional hollow conductor waveguide filled with a material exhibiting the coexistence of chiral magnetic and anomalous quantum Hall effects supports the propagation of transverse electromagnetic modes. This simple setup provides a direct and optically feasible method to probe the simultaneous presence of these phenomena, potentially enabling the detection of an emergent axion-like field within condensed matter systems.
\end{abstract}

\maketitle

%%%%%%%%%%%%%%%%%%%%%%%%%%%%%%%%%%%%%%
%%%%%%%%%%%%%%%%%%%%%%%%%%%%%%%%%%%%%%
%\section{Introduction}
%%%%%%%%%%%%%%%%%%%%%%%%%%%%%%%%%%%%%%
%%%%%%%%%%%%%%%%%%%%%%%%%%%%%%%%%%%%%%

\textit{Introduction.} Recently, the chiral magnetic effect (CME) and the anomalous quantum Hall effect (AQHE) have been predicted to occur simultaneously in antiferromagnetic (AFM) insulators with spin-orbit coupling (SOC), with fluctuations of their order parameter (Néel field) linked to an emergent dynamical axion field \cite{Sekine-PRL116-2016} (for a review see Ref.~\cite{Sekine-JAP-2021}). Indeed, an analogue of such axion field has been previously predicted to occur in topological magnetic insulators. There, magnetic fluctuations couple to the electromagnetic field likewise the axionic field does \cite{Zhang-Nature-Physics-2010}. In addition, Weyl semimetals (WSMs) with broken time reversal and space reflection symmetries could also host the coexistence of both effects \cite{Burkov-PRB85-165110-2012,Chang-etal-PRB97-041104-2018,Heidari-PRB106-195148-2022}.

The AQHE arises when time-reversal symmetry is broken by SOC effects, typically observed in the ferromagnetic phase of certain materials. Additionally, topologically protected charged edge states lead to quantized Hall conductivity (for a review, see Ref.~\cite{Nagaosa-RMP-2010}). In turn, CME emerges as a consequence of chiral anomaly yielding an imbalance of right- and left-handed chiral fermions under a magnetic field. In addition, this imbalance is linked to the topology of gauge fields by the Atiyah-Singer index theorem, ensuring that the resulting chiral current is dissipationless  \cite{kharzeev-PRD-2008,kharzeev}. Evidence of such an effect in the quark-gluon plasma have been reported in Refs.~ \cite{Alice-Coll-NPA-2014,Star-Coll-PRL-2015}. More recently, its detection has also been linked to dark matter phenomena \cite{Hong-PRD-110-055036-2024}, where axionic dark particles are predicted to induce alternating currents along magnetic fields in stars and galaxies. 
Although originally proposed in the high energy physics realm, this effect has also recently been investigated in condensed matter systems, so that electric conductivity scaling with the applied magnetic field squared, $|\bbold_\text{ext}|^2$, has been interpreted as a signature of the CME \cite{science.aac6089,TaAs,Kharzeev-Nat-Phys-2016,GdPtBi}, even though its unambiguous identification remains under debate \cite{TaP,PhysRevX.8.031002,NatureReview}. To address these challenges alternative approaches, including optical probing, have been suggested to detect the CME in a more direct and unambiguous way \cite{Ronald-PRB110-174427-2024} (see also the ``Optical experiments'' section in Ref.~\cite{NatureReview} for a brief review).

In turn, waveguide devices are widely used to control the profile, the intensity, and the spot of electromagnetic waves designed to characterize a given sample, to send and receive information, and so forth. For that, transverse electromagnetic (TEM) modes appear to be the most suitable ones for several applications since they ensure high efficiency and minimize losses and dispersion along the guide. However, pure TEM waves are idealized modes which would propagate only under ideal circumstances of both media (vacuum or another linear, isotropic, and homogeneous medium) and boundary conditions (perfect conductor). In practice, due to the finite conductivity of real conductors the best one can do is using suitable materials and geometries that allow almost perfect TEM waves to flow. 

Let us recall that TEM modes are not allowed to propagate in any guide composed of a single conductor, as stemming from limitations of Laplace’s equation in bounded geometries without sources \cite{jackson,griffiths}. However, this may be bypassed provided that one departs from usual Maxwell's electrodynamics \cite{PRB93-045022-2016,Filipini-PRB109-235108-2024}. This can be achieved, for example, by extending Maxwell’s theory to include axion-like electrodynamics \cite{Wilczek}, which has been widely used to describe the topological magnetoelectric effect coming about whenever light interacts with topological insulators, Weyl semimetals, and other peculiar materials \cite{RevModPhys-2018}. Other recent proposals to achieve TEM modes flow include all-dielectric coaxial cables made from omnidirectional reflecting mirrors  \cite{Ibanescu-Science-2000} and metamaterials with extremely large anisotropies  \cite{Catrysse-PRL-2011}. 

In this article, we propose a simple and feasible apparatus designed to optically probe the coexistence of CME and AQHE under the action of a light beam. Indeed, we show that a hollow conductor tube filled with a material supporting both effects allows the propagation of TEM modes in its interior, otherwise forbidden in conventional media. The observation of TEM under these conditions constitutes direct evidence of an emergent axion-like field in the condensed matter realm by purely optical means, avoiding intricate setups like those typically used in electronic transport and related measurements. 

%%%%%%%%%%%%%%%%%%%%%%%%%%%%%%%%%%%%%%%%%%%%%%%
%%%%%%%%%%%%%%%%%%%%%%%%%%%%%%%%%%%%%%%%%%%%%%
%\section{Model and conditions for TEM modes}
%%%%%%%%%%%%%%%%%%%%%%%%%%%%%%%%%%%%%%%%%%%%%%%
%%%%%%%%%%%%%%%%%%%%%%%%%%%%%%%%%%%%%%%%%%%%%%

\textit{The model.} Let us start off by augmenting Maxwell's electrodynamics with the inclusion of the so-called axion term  \cite{Zhang-Nature-Physics-2010,Sekine-JAP-2021,Wilczek,RevModPhys-2018}:
\be\label{action}
S_\theta = \sqrt{\frac{\varepsilon}{\mu}} \frac{\alpha}{\pi} \int dt\, d^3x \,\theta(\bx,t) \ebold \cdot \bbold,
\ee
where $\alpha=e^2/4\pi\varepsilon\hbar v$ is the material fine structure constant and $v=1/\sqrt{\mu\varepsilon}$. Whenever it comes from a constant background 4-vector $\mathfrak{b}^{\mu}=(\mathfrak{b}_0,\bs{\mathfrak{b}})$ coupled to spacetime coordinates, like 
\be\label{axion-field}
\theta(\bx,t) = 2(\bs{\mathfrak{b}} \cdot \bx - \mathfrak{b}_0 t),
\ee
one has a non-dynamical axion field, $\theta$. Since $\mathfrak{b}^{\mu}$ is a constant 4-vector it explicitly breaks Lorentz invariance; however, it respects gauge symmetry so that electric charge is conserved. For simplicity of notation, from now on we define $b^\mu = (2\alpha/\pi)\mathfrak{b}^\mu$. 

By virtue of the axion term, Gauss' law acquires an extra charge density,
\be\label{extracharge}
\frac{\rho_\theta}{\varepsilon} = v \bb \cdot \bbold,
\ee
while Amp\`ere-Maxwell's law is augmented by an extra current, 
\be\label{extracurrent}
\mu\mbf{J}_\theta = \frac{1}{v}(b_0 \mbf{B} - \bb \times \mbf{E}).
\ee
Even though originally proposed as a model for probing novel phenomena concerning violations of Lorentz symmetry \cite{jackiw,cfj,PLB-Unitarity-MCFJ,EPJP-Boundaries-MCFJ,Large-LV-WS}, presumably originated at the very high-energy regime of the primordial universe, such an electrodynamics has also been used as an effective description of exotic states of matter \cite{Zhang-Nature-Physics-2010,Sekine-JAP-2021,RevModPhys-2018,Deng-PRB104-075202-2021}. More specifically, $b_0$-term describes the (static) CME \cite{kharzeev} while $\bb$-terms are associated to AHE \cite{Nagaosa-RMP-2010,chiralmatter}. For instance, in certain AFM compounds with strong SOC effects bringing about space-time variations for the order parameter (Néel field), $\mbf{n}(\bx,t)$, the local fluctuations of the magnetization are regarded as the realization of an emergent (dynamical) axion field, i.e., $\delta \mbf{n}(\bx,t) \propto \delta\theta(\bx,t)$  \cite{Sekine-PRL116-2016,Sekine-JAP-2021}, which may be encompassed by Eq.~\xref{axion-field} in the linear response regime. In turn, for WSMs $2\hbar b_0$ represents the shift in energy (also proportional to the chemical potential difference) of two Weyl nodes in a $P$-breaking WSM while $2\bb$ is the separation distance in reciprocal space of two Weyl nodes in a $T$-breaking WSM  \cite{Burkov-PRB85-165110-2012,Chang-etal-PRB97-041104-2018,Heidari-PRB106-195148-2022}.

In the electromagnetic response of a material exhibiting the AQHE and CME, the AQHE induces gyrotropy, leading to off-diagonal components in the permittivity tensor, while the CME introduces a longitudinal magnetoelectric coupling (see Ref.~\cite{elight} for a review). However, the diagonal components of the permittivity remain isotropic unless additional symmetry-breaking mechanisms, such as Weyl cone tilting in WSMs \cite{PhysRevB.96.195157} or crystal anisotropies, are present. Here, we model the bulk medium as a material with isotropic permittivity and incorporate the anisotropic effects arising from the CME and AQHE using axion electrodynamics, effectively implementing bi-isotropic constitutive relations \cite{elight,PhysRevA.106.042205}. For WSMs, this approach describes the linear response of the system under light beam for negligible cone tilting. For small tilts (type-I WSM) aligned with the waveguide axis (resulting in a uniaxial medium), our approach remains valid, as the TEM modes considered here decouple from longitudinal tilting effects.

%%%%%%%%%%%%%%%%%%%%%%%%%%%%%%%%%%%%%%%%%%%%%%%%%%%%%%%%%
\begin{figure}[!ht] \label{fig:fig1}
\includegraphics[width=250 pt]{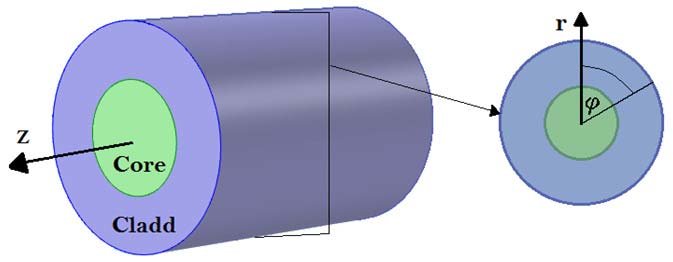}
\caption{\small Schematics of a cylindrical waveguide consisting of a hollow metallic wall (gray cladd) filled with a compound supporting the coexistence of CME and AQHE (the core depicted in light green color). A light beam is imparted to the compound and TEM modes propagate through it when an external magnetic field $\bbold_\text{ext}=B_0 \z$ is turned on.}
\label{fig:schematics}
\end{figure}
%%%%%%%%%%%%%%%%%%%%%%%%%%%%%%%%%%%%%%%%%%%

\textit{Conditions for TEM modes.} Now, we shall establish the conditions which allow TEM modes to propagate through the interior of hollow metallic guides filled with a material supporting AQHE and CME concomitantly, and which are linked to an emergent axion-like field as prescribed by Eq.~\xref{axion-field}. Fig.~\ref{fig:schematics} depicts the specific case of a circular cylindrical guide. The incident light beam has frequency $\omega$ lying within the transparency band gap of the core material, i.e., $\omega_p<\omega<\omega_F$, where $\omega_p$ and $\omega_F$ are the plasma and Fermi frequencies of the material (further details below). A general TEM sinusoidal wave propagating along the $\z$-direction is given by
\begin{align}\label{ansatz}
& \ebold(\bx,t) = \ep(\bx_\bot) e^{i(kz-\omega t)}
\quad \text{and} \quad \ep\cdot\z = 0,
\\[5pt]
& \bbold(\bx,t) = \bp(\bx_\bot) e^{i(kz-\omega t)}
\quad \text{and} \quad \bp\cdot\z = 0,
\label{ansatzB}
\end{align}
where $\bx_\bot=(x,y) = (r\cos\varphi,r\sin\varphi)$ in Cartesian and cylindrical coordinates, respectively. The guide is assumed to be a perfect conductor, so that the homogeneous boundary conditions read \footnote{Indeed, finite conductivity does not jeopardize our results concerning TEM modes since it only imparts usual sources and dissipation, as discussed in textbooks  \cite{jackson,griffiths}.}
\be\label{BCs}
\hat{\mbf{n}} \times \ep|_\mathcal{S} = 0,
\qquad
\hat{\mbf{n}} \cdot \bp|_\mathcal{S} = 0,
\ee
where $\hat{\mbf{n}}$ is the unity normal vector at the guide walls $\mathcal{S}$.

In turn, inside the guide filled with the suitable material, Faraday's and Amp\`ere-Maxwell's laws, augmented by the current arising from the axionic action, Eq.~(\ref{extracurrent}), in the linear response regime, yield \footnote{Strictly speaking, Eq.~\xref{ik1-tem} provides an accurate approximation when spatial variations in the material's magnetization $\mathbf{M}$ are negligible ($\nabla\times\mathbf{M} \approx 0$), and the response to electric fields is linear. Under these conditions, the combined free, polarization, and magnetization currents result in $\mathbf{J} = \left[\sigma - i\omega\varepsilon_0(\varepsilon_r - 1)\right]\mathbf{E}$. Substituting this into Ampère-Maxwell’s equation leads to Eq.~\xref{ik1-tem}, given the definition of $\epsilon_r^*(\omega)$.}
\begin{align}
    & ik\ep - i\omega \bp \times\z = 0,
    \label{ik2-tem}
    \\[5pt]
    &ik\bp + i\frac{\omega}{c^2} \mu_r\epsilon_r^* \ep\times\z = \frac{1}{v} (b_0 \bp\times\z - b_z\ep),
    \label{ik1-tem}
\end{align}
with $\mu_r\equiv \mu/\mu_0$, and where the complex relative permittivity is defined as $\epsilon_r^*(\omega) \equiv \varepsilon_r(\omega) + i\sigma(\omega)/\varepsilon_0\omega$, with $\varepsilon_r \equiv \varepsilon/\varepsilon_0$ denoting its real part and $\sigma$ the material conductivity \cite{ashcroft}. Mutual consistency of Eqs.~\xref{ik2-tem} and \xref{ik1-tem} requires
\be\label{consistency-condition}
\omega = \frac{ck}{\sqrt{\mu_r\epsilon_r^*(\omega)}}
\qquad \text{and} \qquad
b_0 = \frac{\omega}{k} b_z.
\ee

Since we shall obtain transmitted power by TEM waves throughout the guides, let us focus our analysis on the optically transparent regime of the material, i.e., $\omega_p < \omega < \omega_F$. The plasma frequency, $\omega_p$, defines the threshold below which reflection dominates due to collective oscillations of free electrons (plasmons), whereas the Fermi frequency, $\omega_F = E_F/\hbar$, represents the energy scale associated with the Fermi energy. Indeed, whenever $\omega < \omega_F$, the single-band (Drude-like) term dominates, and interband transitions are minimal, preserving transparency. Assuming the validity of Drude approximation, $\epsilon_r^*(\omega)\approx\varepsilon_r(1-\omega^2_p/\omega^2)$ \cite{PRB-Kotov-Lozovik-2016,PRB-Kotov-Lozovik-2018}, so that $ck \gg \omega_p$. Then, the first condition in Eqs.~\xref{consistency-condition} yields a single propagating TEM mode with frequency
\be\label{dispersion-relation}
\omega = \frac{ck}{\sqrt{\mu_r\varepsilon_r}} + \mathcal{O}\left(\frac{\omega_p^2}{c^2k^2}\right) \approx \frac{k}{\sqrt{\mu\varepsilon}} = vk,
\ee
so that light propagates throughout the bulk without dispersion and at maximum speed, $v=1/\sqrt{\mu\varepsilon}$. Notice also that all the frequencies within the medium optical transparency window are allowed to propagate. The second condition in \xref{consistency-condition} is automatically satisfied,
\be\label{b0bz}
b_0 = v b_z,
\ee
since $\omega/k\approx v$ is constant.

Equations \xref{dispersion-relation} and \xref{b0bz} are necessary for TEM mode propagation in this setup, enforcing the consistency between Faraday’s and Ampère-Maxwell’s laws for transverse fields. In a guide filled with a linear, isotropic, and homogeneous medium, such consistency requires dispersionless TEM propagation ($\omega = vk$) \cite{griffiths,jackson}. Our results extend this condition to materials exhibiting axion electrodynamics, where TEM propagation further imposes $b_0 = v b_z$. In this regime, Eq.~\xref{ik2-tem} implies that the electric and magnetic fields must be orthogonal to each other and to the propagation direction within the waveguide as it should be for a TEM mode:
\be\label{fields-tem}
\ep = v \bp \times \z.
\ee

Before proceeding further, we should discuss the feasibility of the condition $b_0 = \omega b_z/k= v b_z$. Recall that in WSMs, $2\hbar b_0$ is the energy gap ($\sim$ chemical potential difference) while $2b_z$ represents the shift in momentum space (along $z$) between two Weyl nodes. Therefore, to satisfy the condition above, tuning of one or both parameters may be in order. This can be accomplished in practice with applied magnetization/magnetic field, pressure and so forth (see, for instance, Refs.~\cite{PRResearch-1-032044R-2019,Nat-Comm-15-1467-2024}). Indeed, even huge variations in $b$-parameters could be achieved by suitable thermal fluctuations, at least in a specific class of WSMs like the compound Cd\textsubscript{2}Re\textsubscript{2}O\textsubscript{7}, as reported in Ref.~\cite{PRB-105-L081117-2022}.

%%%%%%%%%%%%%%%%%%%%%%%%%%%%%%%%%%%%%%%%%
%%%%%%%%%%%%%%%%%%%%%%%%%%%%%%%%%%%%%%%%%
%\section{Results and Discussion}
%%%%%%%%%%%%%%%%%%%%%%%%%%%%%%%%%%%%%%%%%
%%%%%%%%%%%%%%%%%%%%%%%%%%%%%%%%%%%%%%%%%

\textit{Results and Discussion.} In the following, we explicitly discuss three waveguide geometries that support TEM modes under the application of a uniform magnetic field, $\bbold_\text{ext} = B_0 \z$, as long as the guide is filled by a core material with $b_0 = v b_z$. In this situation, Maxwell's equations now read:
\begin{align} 
    & \divbot \ep = \bbot\cdot(v\bp) + b_0 B_0,
    && \nablab \times \ep = 0, 
    \label{gauss-faraday}
    \\[10pt]
    & \nablab\times(v\bp) = b_0B_0\z - \bbot\!\times\ep,
    && \divbot \bp = 0.  
    \label{ampere-div}
\end{align}
Physically, $\bbold_\text{ext} = B_0 \z$ couples to the emergent axion field $\theta$ through its first derivatives ($b_0\propto\partial_t\theta$ and $b_z\propto\partial_z\theta$) and yields effective source terms ($\rho\propto b_0B_0$ and $J_z\propto b_zB_0$), circumventing the prohibition of TEM modes in hollow metallic guides. In single-walled guides, like in cylindrical and rectangular  geometries, one also needs to set  $\bbot \cdot \bp = 0$ to ensure TEM modes, whereas in a slab-type guide this condition is not required.

%%%%%%%%%%%%%%%%%%%%%%%%%%%%%%%%%%%%%%%%%%%%
%%%%%%%%%%%%%%%%%%%%%%%%%%%%%%%%%%%%%%%%%%%%
%\subsection{Cylindrical waveguide}
%%%%%%%%%%%%%%%%%%%%%%%%%%%%%%%%%%%%%%%%%%%%
%%%%%%%%%%%%%%%%%%%%%%%%%%%%%%%%%%%%%%%%%%%%

\textit{Cylindrical waveguide.} The homogeneous Maxwell's equations yield the boundary conditions \xref{BCs}, specifically $\s \times \ep = 0$ and $\s \cdot \bp = 0$ at the walls of a cylinder of radius $R$. Taking $\bbot \cdot \bp = 0$ into account, the non-homogeneous equations become analogous to those describing electrostatic and magnetostatic fields inside a wire with uniform charge and current densities given by $\rho/\varepsilon = b_0 B_0$ and $\mu \mbf{J} = (b_0 B_0/v)\z$, respectively. The solutions are readily obtained as $\ep = (\rho r/2\varepsilon)\s$ and $\bp = (\mu J r/2)\bphi$ along the transverse coordinates $(r, \varphi)$ \cite{griffiths}\footnote{The fields $\ebold$ and $\bbold$ can equivalently be obtained by solving Maxwell’s equations for the electric potential $\Phi$, followed by using $\ep = v \bp \times \z$, as done for other waveguide geometries in the following sections. We chose not to do so to emphasize that the coupling of the applied magnetic to the emergent axion field mimics effective charge and current sources.}. Augmenting them with the propagating factor along $z$, one obtains the field solutions propagating inside the waveguide as 
\begin{align}
\ebold(\bx,t) & = \frac{1}{2} b_0 B_0\, r \cos(kz-\omega t) \s,
\label{E-cylindrical}
\\[5pt]
\bbold(\bx,t) & = \frac{1}{2v} b_0 B_0\, r \cos(kz-\omega t) \bphi + B_0\z,
\label{B-cylindrical}
\end{align}
whereas $\bb=b_r\s+(b_0/v)\z$. These solutions, while unaffected by the homogeneous boundary conditions \xref{BCs}, determine the induced charge and current densities at the guide wall via the inhomogeneous Maxwell equations.

The radiation power density ${\cal P}_\text{rad}=\mbf{S}\cdot\z = \frac{1}{\mu} (\ebold\times\bbold)\cdot\z = \frac{1}{4\mu v} (b_0 B_0)^2 r^2 \cos^2(kz-\omega t)$ yields the time-averaged transmitted power through the entire guide:
\begin{equation}\label{power-cylindric}
P= \int \big<{\cal P}_{\rm rad}\big>_t\, dA=\frac{\pi R^4}{16\mu v}\big(b_0 B_0\big)^2.
\end{equation}
It is noteworthy that such an intensity is also proportional to $(B_0)^2$ and comes from topological grounds brought about by the chiral anomaly according to action \xref{action}, likewise the CME chiral conductivity reported in Ref.~\cite{Kharzeev-Nat-Phys-2016}. Its optical probing, as predicted above, provides an unambiguous and direct signature of the emergent axion field realization by the material filling the waveguide.

%%%%%%%%%%%%%%%%%%%%%%%%%%%%%%%%%%%%%%%%%%%%
%%%%%%%%%%%%%%%%%%%%%%%%%%%%%%%%%%%%%%%%%%%%
%\subsection{Rectangular waveguide}
%%%%%%%%%%%%%%%%%%%%%%%%%%%%%%%%%%%%%%%%%%%%
%%%%%%%%%%%%%%%%%%%%%%%%%%%%%%%%%%%%%%%%%%%%

\begin{figure}[!htb]
\includegraphics[width=1\linewidth]{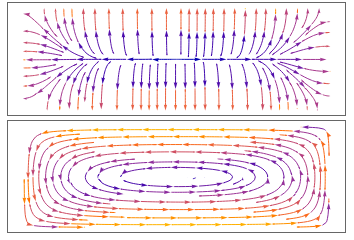}
\caption{Cross-sectional profiles of the electric (top) and magnetic (bottom) fields of a propagating TEM wave inside a rectangular waveguide with a width-to-height ratio of 3:1.}
    \label{fig:rect-external-field}
\end{figure}

\textit{Rectangular waveguide.} The case of a rectangular waveguide with width $w$ and height $h$ can be readily solved using an electrostatic potential $\Phi(x,y)$ that vanishes at the guide walls. By setting $\bs{\mathcal{E}} = - \nablab \Phi$ and $\bbot \cdot \bp = 0$, the nonhomogeneous Maxwell's equations reduce to
\be\label{iii-phi}
\nablab^2 \Phi(x,y) = - b_0 B_0.
\ee
This equation is analogous to the Saint-Venant torsion problem in solid mechanics of a bar with a constant cross-section subjected to torsional loads, where $\Phi$ plays the role of Prandtl's stress function, and the source term is $-2Gk$ with $G$ the shear modulus and $k$ the (constant) twist rate  \cite{timoshenko-goodier}. The solution for the rectangular cross-section is well-known in the literature  \cite{prandtl} and it is
\begin{align} \label{Phi-retangular}
\Phi(x,y) = \sum_{n=1,3,\dots} &\frac{8b_0 B_0 w^2}{(n\pi)^3} \sin\frac{n\pi x}{w} \Bigg[ \cosh\frac{n\pi y}{w} - 1 
\nonumber\\[5pt]
& + \left( \frac{1-\cosh \frac{n\pi h}{w}}{ \sinh \frac{n\pi h}{w} } \right) \sinh\frac{n\pi y}{w} \Bigg].
\end{align}
The corresponding expressions for $\ebold$ and $\bbold=\z\times\ebold/v$ are not very revealing. Figure \ref{fig:rect-external-field} depicts their spatial behavior in a cross-section of the guide (considering the first 20 terms of the summation above). The field lines reveal an effective line-like charge and current densities located along the guide axis (center of the images) due to the coupling between the applied axial magnetic field $\bbold_\text{ext}=B_0\z$ and the component $b_z=b_0/v$, whereas $\bb_\bot$ is everywhere perpendicular to $\bbold$. Of course, the transmitted power through the guide scales with $(B_0)^2$, as in the previous geometry.

%%%%%%%%%%%%%%%%%%%%%%%%%%%%%%%%%%%%%%%%%%%%%%%%%%%
%%%%%%%%%%%%%%%%%%%%%%%%%%%%%%%%%%%%%%%%%%%%

%%%%%%%%%%%%%%%%%%%%%%%%%%%%%%%%%%%%%%%%%%%%
%%%%%%%%%%%%%%%%%%%%%%%%%%%%%%%%%%%%%%%%%%%%
%\subsection{Slab waveguide}
%%%%%%%%%%%%%%%%%%%%%%%%%%%%%%%%%%%%%%%%%%%%
%%%%%%%%%%%%%%%%%%%%%%%%%%%%%%%%%%%%%%%%%%%%

\textit{Slab waveguide.} For a slab waveguide, the condition $\bbot\cdot\bp=0$ is no longer necessary. The slab we consider has height $h$, and it is open along the $x$- and $z$-directions. Since the fields are independent of $x$, Gauss' and Faraday's laws \xref{gauss-faraday} simplify to
\be\label{slab-equations}
\partial_y \mathcal{E}_y = b_y \mathcal{E}_x - b_x \mathcal{E}_y + b_0 B_0,
\qquad
\partial_y \mathcal{E}_x = 0,
\ee
once $\boldsymbol{\mathcal{E}} = v \boldsymbol{\mathcal{B}} \times \z$ is used. Since Faraday's law imposes the boundary condition $\mathcal{E}_x|_{y=\{0,h\}} = 0$, it follows that $\mathcal{E}_x = 0$ throughout the slab, so the $b_y$ component does not affect field solutions in this setup. Writing $\mathcal{E}_y(y) = -d\Phi(y)/dy$, Gauss's law then reads:
\be
\frac{d^2\Phi}{dy^2} + b_x \frac{d\Phi}{dy} = - b_0 B_0.
\ee
Setting slab top and bottom walls at the same potential, say, $\Phi(0) = \Phi(h) = 0$, one obtains \footnote{Recall that even an ordinary slab waveguide with walls at different potentials supports TEM mode propagation. This is not the case, however, when the walls are at the same potential.}
\be
\Phi(y) = \left(
\frac{1-e^{-b_x y}}{1-e^{-b_x h}} h - y
\right) \frac{b_0 B_0}{b_x}\,,
\ee
yielding the electric and magnetic fields propagating inside the slab to be
\begin{align}
\ebold(y,z,t) & = - \left(
\frac{b_x h \, e^{-b_x y}}{1-e^{-b_x h}} - 1
\right) \frac{b_0 B_0}{b_x}
\cos(kz-\omega t) \hat{\mbf{y}},
\label{E-slab}
\\[5pt]
\bbold(y,z,t) & =  
\left(
\frac{b_x h\, e^{-b_x y}}{1-e^{-b_x h}} - 1
\right) \frac{b_0 B_0}{v b_x}
\cos(kz-\omega t) \hat{\mbf{x}},
\label{B-slab}
\end{align}
whereas $\bb=b_x \hat{\mathbf{x}} + b_y \hat{\mathbf{y}} + (b_0/v) \mathbf{\hat{z}}$.

Thus, inside this slab waveguide TEM mode flows with ${\cal P}_{\rm rad} = \frac{1}{\mu v} [ b_x h \, e^{-b_x y}/(1-e^{-b_x h}) - 1 ]^2 (b_0 B_0/b_x)^2 \cos^2(kz-\omega t)$, so that the time-averaged transmitted power per unit length along $x$ reads:
\begin{equation}
\frac{dP}{dx}= \frac{h}{2\mu v} \left[ \frac{b_x h}{2} \text{coth}\left( \frac{b_x h}{2} \right) - 1
\right] \left( \frac{b_0 B_0}{b_x} \right)^2.\\[5pt]
\end{equation} 
Besides $b_0$ (and $b_z$, satisfying $b_0=vb_z$), the slab geometry also offers a way to probe the transverse $\bbot$ parameters, namely, that perpendicular to its walls. In addition, notice that $dP/dx$ is analytic at $b_x=0$ (even though the fields are not) and, in this limit, reduces to $dP/dx = (b_0B_0)^2 h^3/24\mu v$ \footnote{It can be checked that this result agrees with that for a rectangular guide of width $w\to\infty$ in the $\bbot\cdot\bp=0$ case. TEM solutions are obtained by solving $d^2\Phi/dy^2 = -b_0 B_0$ subject to $\Phi(0)=\Phi(h)=0$, yielding $\Phi(y) = -b_0 B_0 y(y-h)/2$. The fields are then $\ebold = b_0 B_0 (y-h/2) \cos(kz-\omega t), \hat{\mbf{y}}$ and $\bbold = (\hat{\mbf{z}} \times \ebold)/v$, and the time-averaged transmitted power per width-length is $dP/dx = (b_0B_0)^2 h^3/24\mu v$.}.

In the work of Ref.~\cite{Halterman19}, a slab-type geometry made from ultrathin Weyl semimetal films in the epsilon-near-zero (ENZ) regime is also considered. There, localized TM-like modes with slow energy transport occur. In contrast, we have been analyzing a hollow waveguide uniformly filled with a medium exhibiting CME and AQHE, enabling bulk propagation of TEM modes via fine-tuning of $b_0$ and $b_z$. Thus, distinct findings are attributed to key differences in the geometry (thin film vs.\ hollow waveguide), physical mechanisms (anisotropic ENZ vs.\ axion electrodynamics), and mode types (localized TM vs.\ propagating TEM).

%%%%%%%%%%%%%%%%%%%%%%%%%%%%%%%%%%%%%%%%%%%%%%%%%%%
%%%%%%%%%%%%%%%%%%%%%%%%%%%%%%%%%%%%%%%%%%%%%%%%%%%
%\section{Concluding remarks}
%%%%%%%%%%%%%%%%%%%%%%%%%%%%%%%%%%%%%%%%%%%%%%%%%%%
%%%%%%%%%%%%%%%%%%%%%%%%%%%%%%%%%%%%%%%%%%%%%%%%%%%

\textit{Concluding remarks}. Whenever a usual metallic waveguide is filled with a material encompassing both chiral magnetic and anomalous quantum Hall effects, then TEM modes are shown to propagate throughout its bulk. We illustrate our findings by considering the cases of cylindrical, rectangular, and slab-type geometries. In every case, the guide is immersed in a homogeneous axial magnetic field that couples to the emergent axion-like field. The transmitted power by TEM waves is proportional to $(B_0)^2$, which is a direct and unambiguous signature of both CME and AQHE coexisting in the material.

Experimental observation of these predictions appears to be direct evidence of an emergent axion-like field in the condensed matter realm by purely optical means, avoiding intricate setups like those typically used in electronic transport and related measurements. In addition, once AQHE can also arise from a nonzero integral of Berry curvature, it should be stressed that usual transport measurements alone cannot decisively relate it to an axion-like field  \cite{Deng-PRB104-075202-2021}. Thus, other experimental evidence is mandatory to unambiguously trace both effects back to the chiral anomaly.

%%%%%%%%%%%%%%%%%%%%%%%%%%%%%%%%%%%%%%
%%%%%%%%%%%%%%%%%%%%%%%%%%%%%%%%%%%%%%
%\begin{acknowledgments}
\textit{Acknowledgments.} The authors thank the Brazilian agencies CAPES, CNPq, FAPEMIG, INCT/CNPq -- {\it Spintr\^onica e Nanoestruturas Magn\'eticas Avan\c{c}adas (INCT-SpinNanoMag)}, and {\it Rede Mineira de Nanomagnetismo/FAPEMIG} for financial support. 
%\end{acknowledgments}
%%%%%%%%%%%%%%%%%%%%%%%%%%%%%%%%%%%%%%
%%%%%%%%%%%%%%%%%%%%%%%%%%%%%%%%%%%%%%

\vfill

%%%%%%%%%%%%%%%%%%%%%%%%%%%%%%%%%%%%%%
\bibliography{references}
%%%%%%%%%%%%%%%%%%%%%%%%%%%%%%%%%%%%%%

%%%%%%%%%%%%%%%%%%%%%%%%%%%%%%%%%%%%%%
%%%%%%%%%%%%%%%%%%%%%%%%%%%%%%%%%%%%%%
\end{document}